# A Walk Through the Approximations of Ab Initio Multiple Spawning

Benoit Mignolet[1*] and Basile F. E. Curchod[2*]

[1]*Theoretical Physical Chemistry, UR MolSYS, B6c, University of Liège, B4000, Liège, Belgium*

[2]*Department of Chemistry, Durham University, South Road, Durham DH1 3LE, UK*

*Corresponding authors: bmignolet@uliege.be, basile.f.curchod@durham.ac.uk*

Full Multiple Spawning offers an in principle exact framework for excited-state dynamics, where nuclear wavefunctions in different electronic states are represented by a set of coupled trajectory basis functions that follow classical trajectories. The couplings between trajectory basis functions can be approximated to treat molecular systems, leading to the Ab Initio Multiple Spawning method, which has been successfully employed to study the photochemistry and photophysics of several molecules. However, a detailed investigation of its approximations and their consequences is currently missing in the literature. In this work, we simulate the explicit photoexcitation and subsequent excited-state dynamics of a simple system, LiH, and we analyze (*i*) the effect of the Ab Initio Multiple Spawning approximations on different observables and (*ii*) the convergence of the Ab Initio Multiple Spawning results towards numerically exact quantum dynamics upon a progressive relaxation of these approximations. We show that, despite the crude character of the approximations underlying Ab Initio Multiple Spawning for this low-dimensional system, the qualitative excited-state dynamics is adequately captured, and affordable corrections can further be applied to ameliorate the coupling between trajectory basis functions.

## I. Introduction

Describing the dynamics of a molecule in its excited electronic states, when the Born-Oppenheimer approximation breaks down, is of great importance to understand light-triggered phenomena. Besides the evident electronic structure problem, such nonadiabatic dynamics constitute a stringent challenge for theoretical chemistry due to the presence of important nuclear quantum effects in the excited-state dynamics.[1-3] A full quantum propagation of the nuclear degrees of freedom is computationally affordable only for small molecules or molecular systems whose excited-state dynamics can be described by a reduced number of nuclear degrees of freedom.[4,5] Hence, a plethora of methodologies have been proposed over the last decades to approximate the quantum nonadiabatic molecular dynamics of larger molecules in their full configuration space: trajectory surface hopping,[6,7] semiclassical approaches,[8,9] quantum-classical Liouville approaches,[10-13] symmetrical quasi-classical windowing,[14-18] linearized nonadiabatic dynamics,[19] Bohmian dynamics,[20-25] or exact-factorization based mixed quantum/classical algorithms.[26-30] In the following, we focus on a particular subset of nonadiabatic techniques that proposes to describe nuclear wavefunctions as a linear combination of travelling Gaussian basis function, called trajectory basis functions (TBFs). The swarm of TBFs can be seen as a moving grid that will follow the nuclear wavepackets in the nonadiabatic dynamics, ensuring a proper description of amplitude transfer in regions of strong nonadiabatic coupling. The idea of using Gaussian functions for quantum dynamics emerged with the seminal work by Heller,[31-33] and different methods have since then emerged for nonadiabatic dynamics, differing in the way they propagate the TBFs.[34-40] In Full Multiple Spawning (FMS), the TBFs follow classical trajectories and their number can increase during the dynamics thanks to a spawning algorithm, ensuring an adequate description of nonadiabatic processes.[39-42] The FMS framework is *in principle* exact in the limit where a large number of TBFs is employed. By applying approximations on the coupling between TBFs, FMS becomes the Ab Initio

Multiple Spawning (AIMS), which is compatible with on-the-fly nonadiabatic dynamics, *i.e.*, the electronic structure quantities required for the nuclear propagation does not need to be precomputed but can be calculated at each integration time step.[43, 44] AIMS has become a method of choice for nonadiabatic molecular dynamics and has been successfully applied to a large number of molecular systems.[44-57] Surprisingly though, only a few studies have touched on the implication of some of the approximations in AIMS,[39, 42, 58-60] and a general assessment of these approximations and their potential breakdown is unfortunately lacking.

In this work, we aim at filling this gap by offering a comprehensible test of all the approximations connecting AIMS to FMS, as well as the basis set convergence of FMS, using a numerically exact solution of the time-dependent Schrödinger equation as reference. Testing the different approximations of AIMS in the context of nonadiabatic dynamics might appear challenging at first glance, as their effects are likely to strongly differ depending on the type of nonadiabatic transitions (single *vs* multiple nonadiabatic passage, for example[7, 61]). In the following, we instead capitalize on the recently developed XFFMS framework, which explicitly incorporates the coupling with an external electromagnetic field in FMS (and AIMS). In this way, one can selectively produce a superposition of nuclear wavepackets, observe their decoherence, and probe their interaction at a later time, both from an interference perspective but also by applying a second pulse and measure variations in electronic populations. Hence, this formalism will allow us to strongly challenge the different approximations of (XF)AIMS and (XF)FMS.

## II. Theory

### IIa. Nonadiabatic Molecular Quantum Dynamics

The central goal of any nonadiabatic methods is to propose the most efficient and accurate approximation to the time-dependent molecular Schrödinger equation

$$i\frac{\partial \Psi(\mathbf{r},\mathbf{R},t)}{\partial t} = \hat{H}(\mathbf{r},\mathbf{R})\Psi(\mathbf{r},\mathbf{R},t), \qquad (1)$$

where $\mathbf{r}$ and $\mathbf{R}$ denote collective variables for the electronic and nuclear coordinates (we use atomic units throughout this article). The molecular Hamiltonian is given by

$$\begin{aligned}\hat{H}(\mathbf{r},\mathbf{R}) &= \hat{T}_{nuc} + \hat{T}_e + V_{e-e}(\mathbf{r}) + V_{e-n}(\mathbf{r},\mathbf{R}) + V_{n-n}(\mathbf{R}) + V_{ext}(\mathbf{r},\mathbf{R},t) \\ &= \hat{T}_{nuc} + \hat{H}_{el}(\mathbf{r},\mathbf{R}) + V_{ext}(\mathbf{r},\mathbf{R},t)\end{aligned} \qquad (2)$$

and contains the kinetic energy and interaction potential operators for both electrons and nuclei as well as an interaction potential between the molecule and an external electromagnetic field $\underline{\mathbf{E}}(t)$, defined as $V_{ext}(\mathbf{r},\mathbf{R},t) = -\underline{\boldsymbol{\mu}}(\mathbf{r},\mathbf{R}) \cdot \underline{\mathbf{E}}(t)$ with $\underline{\boldsymbol{\mu}}(\mathbf{r},\mathbf{R}) = \underline{\boldsymbol{\mu}}^e(\mathbf{r}) + \underline{\boldsymbol{\mu}}^n(\mathbf{R})$ (underlined bold symbols highlights 3D vectors).

Solving the time-independent Schrödinger equation with the electronic Hamiltonian $\hat{H}_{el}(\mathbf{r},\mathbf{R})$ at a fixed nuclear configuration provides electronic eigenstates $\Phi_J(\mathbf{r};\mathbf{R})$, $\hat{H}_{el}(\mathbf{r},\mathbf{R})\Phi_J(\mathbf{r};\mathbf{R}) = E_J^{el}(\mathbf{R})\Phi_J(\mathbf{r};\mathbf{R})$. The solutions of this electronic Schrödinger equation can be used as a basis to express the electronic degrees of freedom, here in the adiabatic representation, leading to the so-called Born-Huang representation of the total molecular wavefunction:[62, 63]

$$\Psi(\mathbf{r},\mathbf{R},t) = \sum_{J}^{\infty} \Omega_J(\mathbf{R},t)\Phi_J(\mathbf{r};\mathbf{R}). \qquad (3)$$

In Eq. (3), $\Omega_J(\mathbf{R},t)$ represents a nuclear amplitude in electronic state $J$. We note at this stage that the Born-Huang representation is not the only possible one and recent works have showed that an exact factorization of the total time-dependent molecular wavefunction is also possible.[64-66]

Upon insertion of Eq.(3) into the time-dependent molecular Schrödinger equation (Eq.(1)) and after some algebra, one can obtain a set of coupled equations of motion for the nuclear amplitudes

$$i\frac{\partial \Omega_I(\mathbf{R},t)}{\partial t} = \left(\hat{T}_N + E_I^{el}(\mathbf{R}) + V_{II}^{ext}(\mathbf{R})\right)\Omega_I(\mathbf{R},t)$$

$$-\sum_J \left[\sum_{\rho=1}^{3N}\left(\frac{1}{M_\rho}\langle\Phi_I|\frac{\partial}{\partial R_\rho}|\Phi_J\rangle_\mathbf{r}\frac{\partial}{\partial R_\rho} + \frac{1}{2M_\rho}\langle\Phi_I|\frac{\partial^2}{\partial R_\rho^2}|\Phi_J\rangle_\mathbf{r}\right) - V_{IJ}^{ext}(\mathbf{R})\right]\Omega_J(\mathbf{R},t), \quad (4)$$

with $V_{IJ}^{ext}(\mathbf{R}) = \langle\Phi_I|V_{ext}|\Phi_J\rangle_\mathbf{r}$. This equation is expressed in the adiabatic representation, for which nonadiabatic coupling vectors (NACVs) – $\mathbf{d}_{IJ}(\mathbf{R}) = \langle\Phi_I|\frac{\partial}{\partial \mathbf{R}}|\Phi_J\rangle_\mathbf{r}$ – and second-order nonadiabatic couplings – $D_{IJ}(\mathbf{R}) = \langle\Phi_I|\frac{\partial^2}{\partial \mathbf{R}^2}|\Phi_J\rangle_\mathbf{r}$ – are non-zero. In the diabatic representation, all nonadiabatic couplings are strictly zero, but the electronic Hamiltonian will no more be diagonal. A way to solve Eq.(4) would consist in expressing it on a grid, a method that we will denote as *numerically exact quantum dynamics* (QD) in the following.[67, 68] In such a representation, the nuclear wavefunction in electronic state $J$ would read

$$\Omega_J(R_1,\ldots,R_f,t) = \sum_{j_1\cdots j_f=1}^{N_1\cdots N_f} C_{j_1\cdots j_f}^{(J)}(t)\prod_{\kappa=1}^{f}\chi_{j_\kappa}^{(\kappa)}(R_\kappa) \quad (5)$$

using a common notation from the literature. Here, $f$ labels a selected nuclear degree of freedom with corresponding number of grid basis functions (points) $N_f$ each labeled by $j_\kappa$. Different types of basis functions were proposed for $\chi_{j_\kappa}^{(\kappa)}(R_\kappa)$ and the interested reader is referred to the literature for more details.[4, 67, 69] While QD offers an accurate description of nuclear quantum effects, its cost obviously grows exponentially with the number of nuclear degrees of freedom. Methods like the Multi-Configuration Time-Dependent Hartree (MCTDH) have been developed to circumvent these limitations and push quantum dynamics towards larger molecular systems.[4, 5, 67, 70]

*IIb. Full Multiple Spawning*

As discussed in the previous section, the nuclear wavefunction amplitudes can be expressed on a grid. One may also choose to represent these nuclear wavefunctions by a linear combination of multidimensional frozen Gaussians, for example. In this case, we can write each nuclear wavefunction as $\Omega_J(\mathbf{R},t) = \sum_i^{N_J} C_i^{(J)}(t) \chi_i^{(J)}(\mathbf{R}; \bar{\mathbf{R}}_i^{(J)}, \bar{\gamma}_i^{(J)}(t), \alpha)$, where $\chi_i^{(J)}(\mathbf{R}; \bar{\mathbf{R}}_i^{(J)}, \bar{\gamma}_i^{(J)}(t), \alpha)$ represents the $i^{th}$ multidimensional Gaussian in electronic state $J$ (note that the electronic state is a label and therefore put in brackets) centered at position $\bar{\mathbf{R}}_i^{(J)}$ with corresponding (complex) amplitude $C_i^{(J)}(t)$, phase $\bar{\gamma}_i^{(J)}(t)$, and frozen width $\alpha$. In the limit where $N_J$ tends towards a large number, *i.e.*, we cover the nuclear configuration space with Gaussian functions, we expect to approach a numerically exact solution of the time-dependent Schrödinger equation, in a formalism that we coin here *Frozen-Gaussian quantum dynamics* (FGQD). However, the scaling problem remains the same as for QD.

One interesting idea was to consider that Gaussian functions may not be fixed but can instead travel over time. In particular, if one finds an adequate dynamics for the Gaussians such that they would follow the dynamics of the nuclear wavefunctions, we can substantially truncate the number of Gaussian functions required for their description. Ideally, we would have a moving grid that always ensures a good support for the nuclear wavefunctions, *i.e.*, Gaussian functions would be found only in regions where the nuclear amplitudes are non-zero. Hence, we would move from static frozen Gaussians $\chi_i^{(J)}(\mathbf{R}; \bar{\mathbf{R}}_i^{(J)}, \bar{\gamma}_i^{(J)}(t), \alpha)$ to trajectory basis functions $\chi_i^{(J)}(\mathbf{R}; \bar{\mathbf{R}}_i^{(J)}(t), \bar{\mathbf{P}}_i^{(J)}(t), \bar{\gamma}_i^{(J)}(t), \alpha)$, whose position $\bar{\mathbf{R}}_i^{(J)}(t)$ and momentum $\bar{\mathbf{P}}_i^{(J)}(t)$ centers can change over time according to some given equations of motion. Different recipes for the propagation of the TBFs in excited-state dynamics were proposed (MultiConfigurational Ehrenfest,[36-38, 71, 72] variational multiconfiguration Gaussian,[34, 35, 73-75] Ab Initio Multiple Cloning,[76, 77] and other more recent schemes[78, 79]) and we will focus here on the *Full Multiple Spawning* (FMS)[39-42] technique.

In FMS, each nuclear wavefunction component in the Born-Huang representation is expanded as a linear combination of frozen Gaussian functions, which follow classical trajectories. The FMS version of the Born-Huang representation therefore reads

$$\Psi(\mathbf{r},\mathbf{R},t) = \sum_{J}^{\infty}\sum_{i}^{N_J} C_i^{(J)}(t) \chi_i^{(J)}\left(\mathbf{R}; \bar{\mathbf{R}}_i^{(J)}(t), \bar{\mathbf{P}}_i^{(J)}(t), \bar{\gamma}_i^{(J)}(t), \alpha\right) \Phi_J(\mathbf{r};\mathbf{R}). \qquad (6)$$

If the TBFs are properly distributed initially and a sufficiently large number of them is used, FMS could *in principle* reach the accuracy of FGQD, while needing a smaller number of Gaussian functions thanks to their time dependence.

An equation of motion for the complex amplitudes $C_j^{(I)}(t)$ can be obtained by inserting Eq. (6) into the molecular time-dependent Schrödinger equation, left-multiplication by $\left(C_k^{(I)}(t)\chi_k^{(I)}\left(\mathbf{R};\bar{\mathbf{R}}_k^{(I)}(t),\bar{\mathbf{P}}_k^{(I)}(t),\bar{\gamma}_k^{(I)}(t),\alpha\right)\Phi_I(\mathbf{r};\mathbf{R})\right)^*$, and integration over both nuclear and electronic coordinates:

$$\frac{d\mathbf{C}^I}{dt} = -i\left(\mathbf{S}_{II}^{-1}\right)\left[\left[\mathbf{H}_{II} - i\dot{\mathbf{S}}_{II}\right]\mathbf{C}^I + \sum_{J\neq I}\mathbf{H}_{IJ}\mathbf{C}^J\right], \qquad (7)$$

where $(\mathbf{S}_{II})_{ki} = \langle \chi_k^{(I)} | \chi_i^{(I)} \rangle_{\mathbf{R}}$ and $(\dot{\mathbf{S}}_{II})_{ki} = \langle \chi_k^{(I)} | \frac{\partial}{\partial t} \chi_i^{(I)} \rangle_{\mathbf{R}}$ are overlap matrices, and $(\mathbf{H}_{IJ})_{ki} = \langle \chi_k^{(I)} \Phi_I | \hat{H} | \chi_i^{(J)} \Phi_J \rangle_{\mathbf{R},\mathbf{r}}$ is an Hamiltonian matrix element in the Gaussian basis. The Hamiltonian matrix couples TBFs together, and an element has the typical form

$$\begin{aligned}H_{ki}^{IJ} &= \langle \chi_k^{(I)} | \hat{T}_{nuc} | \chi_i^{(J)} \rangle_{\mathbf{R}} \delta_{IJ} + \langle \chi_k^{(I)} | E_I^{el} | \chi_i^{(J)} \rangle_{\mathbf{R}} \delta_{IJ} \\ &\quad - \langle \chi_k^{(I)} | \sum_{\rho=1}^{3N} \frac{1}{M_\rho} \langle \Phi_I | \frac{\partial}{\partial R_\rho} | \Phi_J \rangle_{\mathbf{r}} \frac{\partial}{\partial R_\rho} | \chi_i^{(J)} \rangle_{\mathbf{R}} - \langle \chi_k^{(I)} | \sum_{\rho=1}^{3N} \frac{1}{2M_\rho} \langle \Phi_I | \frac{\partial^2}{\partial R_\rho^2} | \Phi_J \rangle_{\mathbf{r}} | \chi_i^{(J)} \rangle_{\mathbf{R}} \\ &\quad - \langle \chi_k^{(I)} \Phi_I | \underline{\mu} | \chi_i^{(J)} \Phi_J \rangle_{\mathbf{R},\mathbf{r}} \cdot \underline{\mathbf{E}}(t)\end{aligned} \qquad (8)$$

Let us describe the different terms in the right-hand side of this equation. The first two terms, related to the nuclear kinetic energy operator and the electronic energy, couple TBFs evolving in the same electronic state. The third term contains the NACVs and couples exclusively TBFs in different electronic states. The fourth term contains the second-order nonadiabatic couplings, which will both contribute an intra- and an interstate coupling between TBFs. We note that these terms are quite often neglected in any practical applications (see Ref.[63, 80-82] for discussions on diagonal and off-diagonal second-order nonadiabatic couplings). These four contributions to the Hamiltonian matrix elements are the original coupling terms in FMS. The last term in Eq. (8) is due to the coupling with an external electromagnetic field $\mathbf{E}(t)$ and provides a new coupling between TBFs with both an intra- and an interstate contributions. FMS with this extra term is called *eXternal Field Full Multiple Spawning* (XFFMS).

As mentioned before, FMS would tend to an exact solution of the time-dependent Schrödinger equation in the limit of a large number of TBFs in each electronic state, *i.e.*, large $N_J$. FMS, however, proposes to replace the large number of TBFs by an algorithm, the *spawning algorithm*, that will dynamically *adapt* the number of TBFs during the entire simulation, increasing the number of basis functions available to describe adequately quantum events such as nonadiabatic or photoexcitation processes. In other words, the spawning algorithm requires that $N_J \to N_J(t)$, and it ensures that the dynamics is carried out with the optimal number of TBFs at any time of the simulation. The spawning algorithm is therefore at the heart of the FMS (and XFFMS) method and suggests a robust way of extending the number of TBFs in the dynamics when needed.[43] Briefly, if a TBF enters a region of strong coupling – due to nonadiabatic effects or an external electromagnetic field – a new TBF will be spawned onto the coupled electronic state(s). After a spawning event, the size of the matrices in Eq.(7) is extended by the corresponding number of newly-created TBFs, ensuring therefore a proper coupling pattern between them, and a physical transfer of nuclear amplitude between the coupled electronic states. For more details on the different spawning algorithms, the reader is referred to previous works.[43, 83, 84]

*IIc. Ab Initio Multiple Spawning*

In the following, we will discuss the two main approximations applied on the FMS framework to obtain the so-called *Ab Initio Multiple Spawning* (AIMS) or *eXternal Field AIMS* (XFAIMS) techniques.[44, 85] The Hamiltonian matrix elements in FMS (Eq.(8)) require an integration over the entire nuclear configuration space. Clearly, such integration is prohibitively expensive to treat molecules, and one is forced to approximate the Hamiltonian matrix elements. Owing to the localized nature of Gaussian functions, the Hamiltonian matrix elements can be approximated by a Taylor expansion centered at the centroid position between two TBFs $\bar{\mathbf{R}}_{ki}^{(IJ)} = \frac{\bar{\mathbf{R}}_{k}^{(I)} + \bar{\mathbf{R}}_{i}^{(J)}}{2}$ (the same applies for TBFs in the same electronic state). Hence, any electronic structure quantity $\vartheta(\mathbf{R})$ can be expressed as

$$\vartheta(\mathbf{R}) = \vartheta(\bar{\mathbf{R}}_{ki}^{(IJ)}) + \sum_{\rho}^{3N} (R_\rho - \bar{R}_{\rho,ki}^{(IJ)}) \frac{\partial \vartheta(\mathbf{R})}{\partial R_\rho}\bigg|_{R_\rho = \bar{R}_{\rho,ki}^{(IJ)}}$$
$$+ \frac{1}{2} \sum_{\rho,\rho'}^{3N} (R_\rho - \bar{R}_{\rho,ki}^{(IJ)}) \frac{\partial^2 \vartheta(\mathbf{R})}{\partial R_\rho \partial R_{\rho'}}\bigg|_{R_\rho = \bar{R}_{\rho,ki}^{(IJ)}, R_{\rho'} = \bar{R}_{\rho',ki}^{(IJ)}} (R_{\rho'} - \bar{R}_{\rho',ki}^{(IJ)}) + \dots \quad . \quad (9)$$

This Taylor expansion can be truncated at different orders. In AIMS, a saddle-point approximation (SPA) of order zero is applied, meaning that any electronic structure quantity – electronic energy, NACVs, or (transition) dipole moments – in the integrals forming the Hamiltonian matrix elements is approximated by $\vartheta(\mathbf{R}) \approx \vartheta(\bar{\mathbf{R}}_{ki}^{(IJ)})$.[39, 40] Within the SPA-0, the integrals in Eq.(8) take the simple form

$$\langle \chi_k^{(I)} | \vartheta | \chi_i^{(J)} \rangle_{\mathbf{R}} \approx \vartheta(\bar{\mathbf{R}}_{ki}^{(IJ)}) \langle \chi_k^{(I)} | \chi_i^{(J)} \rangle_{\mathbf{R}} \quad .$$

The second approximation is linked to the coupling between TBFs at the beginning of the FMS dynamics. In FMS, the initial nuclear wavefunction at time $t_0$ in electronic state $J$, $\Omega_J(\mathbf{R},t_0)$, is represented by a linear combination of $N_J(t_0)$ frozen Gaussians – sometimes called the *parent* TBFs:

$$\Omega_J(\mathbf{R},t_0) = \sum_{i}^{N_J(t_0)} C_i^{(J)}(t_0) \chi_i^{(J)}\left(\mathbf{R}; \overline{\mathbf{R}}_i^{(J)}(t_0), \overline{\mathbf{P}}_i^{(J)}(t_0), \overline{\gamma}_i^{(J)}(t_0), \alpha\right) \tag{10}$$

The set of initial complex coefficients $\left\{C_i^{(J)}(t_0)\right\}_{i=1}^{N_J(t_0)}$ and the position of the Gaussian functions are chosen such that they provide an accurate description of the initial nuclear wavefunction. Hence, the FMS dynamics starts with a group of $N_J(t_0)$ *coupled* parent TBFs. As AIMS is concerned with the dynamics of molecules, *i.e.*, systems with a large number of nuclear degrees of freedom, one would expect that the initial nuclear wavefunction rapidly spreads and, therefore, that the initial parent TBFs rapidly move away from each other, meaning that their mutual coupling rapidly drops to zero. In other words, it would not be a bad approximation to propagate the *parent* TBFs independently already from time $t_0$. In this *independent first generation approximation* (IFGA), the initial conditions – positions and momenta – for *one* parent TBF are simply sampled from a given distribution (often a Wigner distribution) and the complex amplitude for this TBF is set to $C_1^{(J)}(t_0) = 1.0$. The parent TBF is propagated and can spawn new children TBFs – all the TBFs descending from a given parent TBF will be fully coupled. The process is repeated for a large number of parent TBFs, all run independently. The process is continued until convergence and the result of interest is obtained by averaging over all initial conditions *incoherently*.[43, 59] Hence, for a representation of the total molecular wavefunction given by

$\Psi(\mathbf{r},\mathbf{R},t) = \sum_{\beta}^{N_{ini}} \tilde{\Psi}_\beta(\mathbf{r},\mathbf{R},t)$, where $\beta$ labels an initial parent TBF and all its descendants and

$\tilde{\Psi}_\beta(\mathbf{r},\mathbf{R},t) = \sum_{J}^{\infty} \sum_{i}^{N_{J,\beta}} C_{i,\beta}^{(J)}(t) \chi_{i,\beta}^{(J)}\left(\mathbf{R}; \overline{\mathbf{R}}_{i,\beta}^{(J)}(t), \overline{\mathbf{P}}_{i,\beta}^{(J)}(t), \overline{\gamma}_{i,\beta}^{(J)}(t), \alpha\right) \Phi_J(\mathbf{r};\mathbf{R})$, the IFGA implies that there is no coupling between the different $\beta$ branches of the simulation, e.g., $C_{i,\beta}^{(J)}(t)$ is not coupled to $C_{i,\beta'}^{(J)}(t)$, while FMS would couple all TBFs from any $\beta$-branches at any time.[59] (It is important to keep in mind that the spawning algorithm implies that $N_{J,\beta} \to N_{J,\beta}(t)$.)

To summarize, AIMS emerges from FMS by applying both the SPA-0 and the IFGA. Figure 1 summarizes the layers of approximation that separate QD from AIMS. In this article, we propose to study the effect of these approximations for the dynamics of a photoexcited molecule, LiH.

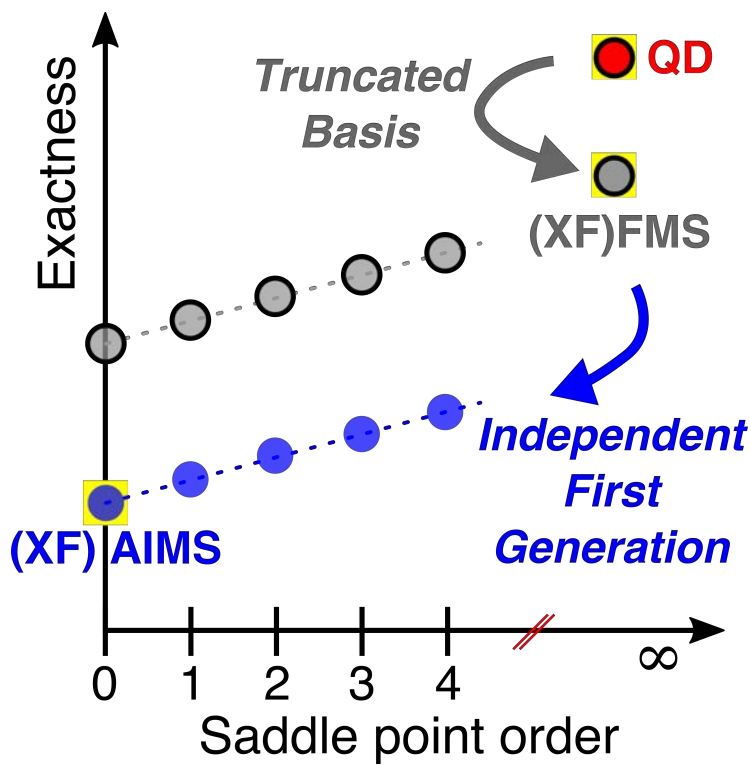

Figure 1: Schematic representation of the different approximations applied to QD to reach (XF)FMS and (XF)AIMS.

**III. Photoexcitation Dynamics Studied and Computational Details**

*IIIa. Photoexcitation Dynamics*

In order to assess and stress the approximations made in (XF)AIMS, we modeled the photoexcitation of the ground state (S$_0$) of the LiH molecule to its lowest excited state (S$_1$). The nuclear dynamics in each state following photoexcitation will strongly differ, owing to the different potential energy curves (PECs) (see Fig. 2). The $S_0$ state exhibits a Morse-like shaped PEC leading to a confined nuclear wavepacket, while the $S_1$ state displays a weakly-bound potential with a larger equilibrium bond length than $S_0$, leading to a $S_1$ nuclear wavepacket that can quickly move towards larger bond distances after photogeneration. Since the $S_1$ PEC is bound, the $S_1$ nuclear wavepacket will eventually come back into the FC region, after a period of approximately 75 fs.[86, 87] Playing with the different dynamics of the nuclear wavepackets in $S_0$ or $S_1$ and the possibility to couple the dynamics with an external laser pulse, we can model specific excitation conditions that will help us test the approximations made in (XF)FMS and (XF)AIMS. In particular, we will concentrate our analysis on the three steps depicted in Fig. 2. First, we will investigate the photoexcitation step (panel A in Fig. 2), focusing on the evolution of the amplitude transfer between the two electronic states during the pulse. Then, the photogenerated nuclear wavepacket on $S_1$ relaxes and quickly leaves the FC region (panel B in Fig. 2) – an event that can be followed by monitoring the time-dependent dipole moment thanks to its interference terms (see below). Finally, the nuclear wavepacket on $S_1$ returns into the FC region and the spatial localization as well as the phase relation between the nuclear wavepackets on $S_1$ and $S_0$ can be probed using a second pulse (panel C in Fig. 2).

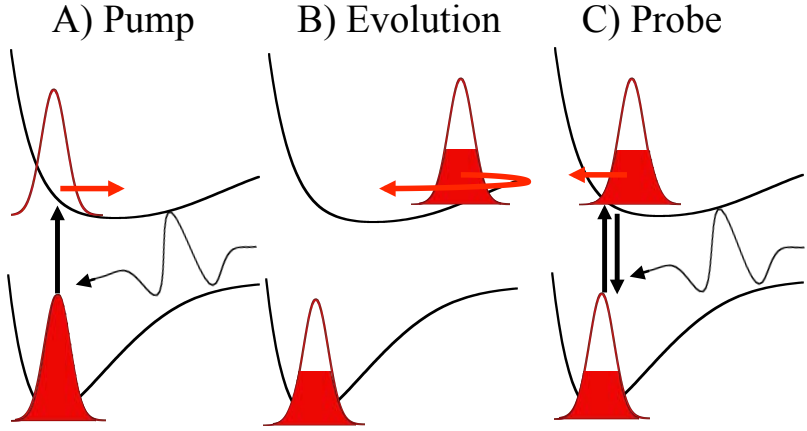

Figure 2: Schematic representation of a pump-probe experiment in LiH in which the system in $S_0$ is photoexcited by a resonant laser pulse to its first excited state $S_1$ (A). The nuclear wavepacket on $S_1$, being on a repulsive part of the PEC, will undergo a dynamics towards longer LiH bond distances until it reflects and comes back into the FC region (B) after 75 fs. This return of the nuclear wavepacket is then probed by a second pulse inducing a transfer of population $S_1 \rightarrow S_0$ or $S_0 \rightarrow S_1$. (The nuclear-wavepacket filling reflects the population in the corresponding electronic state.)

### *IIIb. Computational Details*

In all the simulations presented in this work, the time-dependent electric field of the pulse, $\underline{\mathbf{E}}(t)$, is defined from the derivative of the (Gaussian-shaped) vector potential:

$$\underline{\mathbf{E}}(t) = -\frac{1}{c}\frac{d\underline{\mathbf{A}}(t)}{dt} \qquad (10)$$

with

$$\underline{\mathbf{A}}(t) = \underline{\varepsilon}\frac{cf_0}{\omega}\left[\exp\left[\frac{-(t-t_0)^2}{2\sigma^2}\right]\sin(\omega t + CEP) + \exp\left[\frac{-((t-t_{PP})-t_0)^2}{2\sigma^2}\right]\sin(\omega(t-t_{PP}) + CEP)\right] \qquad (10)$$

where $\underline{\varepsilon}$ is the polarization vector, $c$ the speed of light, $f_0$ is the field strength, and $\sigma$ is related to the pulse duration (the FWHM of the pulse is $2.35\sigma$). $\omega$ corresponds to the carrier frequency and is set to 0.127 a.u. (359 nm) in our simulations, close to resonance with the $S_0$-$S_1$ transition at the $S_0$ equilibrium geometry. The carrier envelope phase (CEP) is the phase difference between the pulse envelope and the oscillation of the electric field. For few cycle pulses, the CEP controls the waveform (sub-femtosecond evolution) of the pulse and can affect the dynamics[88, 89] as well as the branching ratio.[90] The first pulse is centered at the time $t_0$ while the probe pulse is a replica of the pump and is centered at the pump-probe delay time $t_{PP}$.

The exact quantum dynamics grid simulations (QD) are carried out on the first two lowest PECs of LiH, computed at the SA2-CASSCF(4/6)/6-31G level of theory. The $S_0$ PEC has a minimum at 1.66 Å, while the one of the $S_1$ PEC is found at 2.44 Å (Fig. 3). In all simulations, we neglected the excitation to higher excited states as well as the weak nonadiabatic coupling between $S_0$ and $S_1$. The PECs are discretized on a spatial grid with a spacing of 0.02 Bohr from 0.5 to 15.0 Bohr, and the time step for the integration is 0.2 a.u. (0.0048 fs). The initial wavefunction on the ground electronic state has a Gaussian shape and was selected to closely mimics the ground vibrational eigenstate.

The XFFMS and XFAIMS simulations are carried in internal coordinates with a modified version of the AIMS program[91] available in MOLPRO 2012[92], using the very same level of theory as for the QD simulations. XFAIMS and XFFMS mainly differ in the set of initial conditions on $S_0$. For XFFMS, the initial nuclear wavefunction at time $t_0$ is depicted by a set of 9 coupled TBFs with positions varying by step of 0.1 Bohr from 2.73 Bohr (1.44 Å) to 3.53 Bohr (1.87 Å) (see Fig. 3). For XFAIMS, which employs the IFGA (see below), each run is initiated with a single TBF in $S_0$ whose position and momentum are sampled from a Wigner distribution (we note that the sampling of initial conditions is critical as it can influence the overall result[38]). Then, for both XFFMS and XFAIMS, the wavefunction evolves on $S_0$ under the influence of the external electric field described before, until the number of TBFs has to be expanded to properly describe the transfer of amplitude to the $S_1$ state. One or several empty TBFs are spawned on $S_1$ using the spawning algorithm described in Ref. [85]. Only one spawning event occurs at the maximum of the pulse envelope in our simulations. In total, 100 independent

Mignolet and Curchod – Approximations in AIMS – Page 14

XFAIMS runs are performed and incoherently averaged to produce the final result. If there is only one TBF on $S_0$ before the pulse, as it is the case for each individual initial condition with XFAIMS, only one TBF is spawned on $S_1$ (with an nuclear overlap of 1 when the pulse is maximum, *i.e.*, when the transfer of amplitude is expected to be the largest). The situation differs for XFFMS as a result of the larger number of coupled TBFs on $S_0$, which can lead to the spawning of several TBFs on $S_1$. We finally note that the widths employed to form the TBFs (diagonal matrix $\alpha$ in Eq.(6)) is expected to have only a minor effect on population transfer or the time-dependent dipole moment, as discussed in the literature.[93] We tested this fact for the system studied by doubling $\alpha$ from 5 Bohr$^{-2}$ to 10 Bohr$^{-2}$ and did not observe a significant change.

**IV. Results and Discussion**

In the following, we will study how the approximations of XFAIMS affect the description of LiH photoexcitation and its subsequent dynamics. Our discussion will be based on the different steps presented in Fig. 2.

*IVa. Step A: Photoexcitation by a UV femtosecond pulse*

We first investigate the photoexcitation of LiH, originally on $S_0$, by a 0.8 fs laser pulse. At the beginning of the exact QD simulation, the nuclear wavepacket is entirely on $S_0$ (Fig. 3, upper panel, $t = 0$ fs). Then, the laser pulse induces an amplitude transfer between $S_0$ and $S_1$, leading to the creation of a nuclear wavepacket on $S_1$ (Fig. 3, upper panel, 1.2 fs). The $S_1$ wavepacket immediately relaxes and evolves towards longer LiH distance (Fig. 3, upper panel, 9.7-19.4 fs). The time trace of the population on $S_1$, defined as $n_{S_1}(t) = \langle \Omega_{S_1}(t) | \Omega_{S_1}(t) \rangle_\mathbf{R}$, is depicted in Fig. 4 (red line) and shows how the laser pulse transfers population from the ground to the excited electronic state, with a maximum of efficiency when the electric field reaches its maximum.

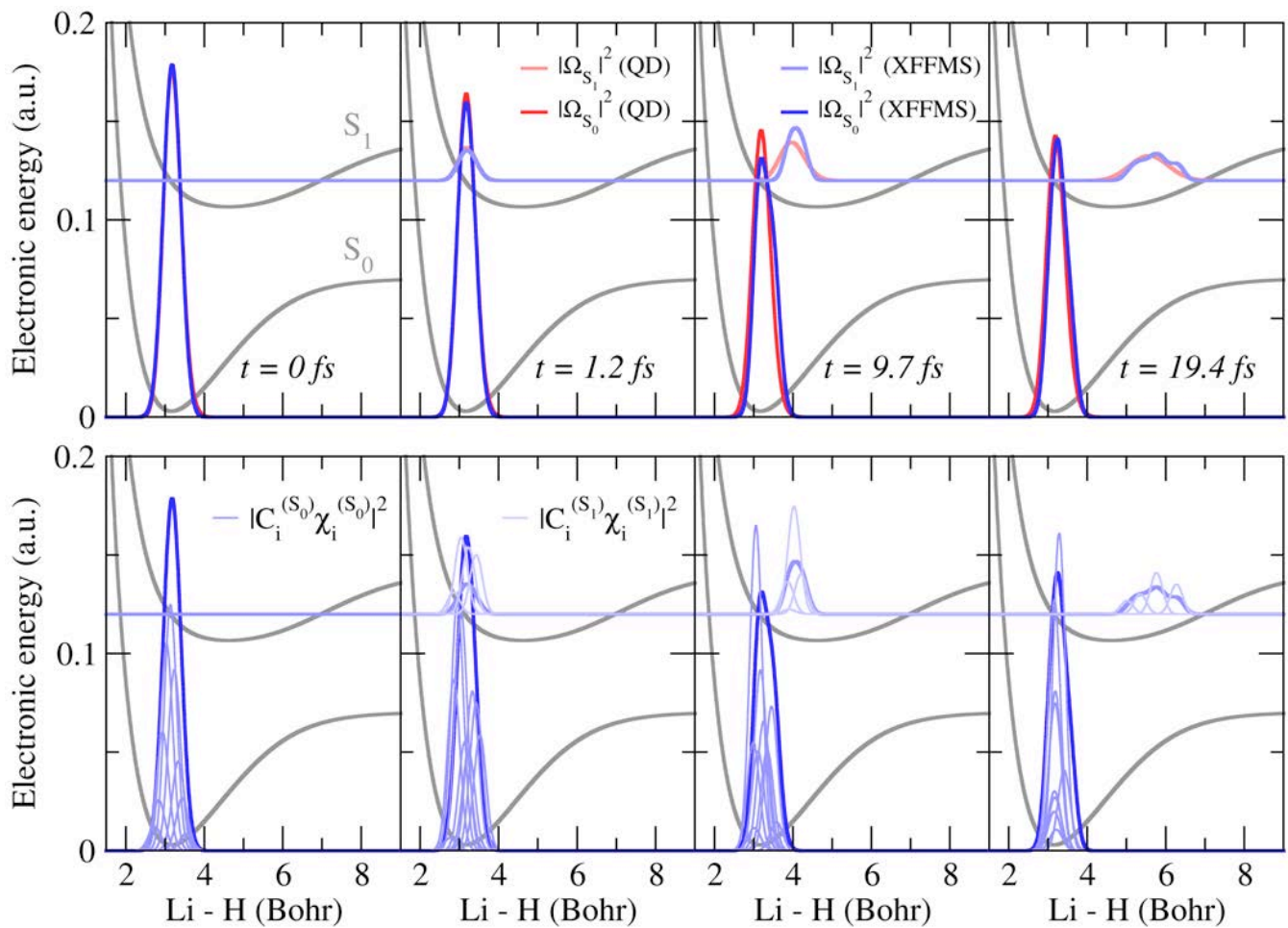

Figure 3: Upper panel: Probability density of the $S_0$ and $S_1$ nuclear wavepackets computed at different times for XFFMS (blue=$S_0$, light blue=$S_1$) and QD (red=$S_0$, light red=$S_1$) dynamics. The $S_0$ and $S_1$ PECs are represented in light gray and computed at the SA2-CASSCF(4/6)/6-31G level of theory. The nuclear wavepackets have been rescaled for visibility. Lower panel: Schematic representation of the TBFs (thin lines) in the XFFMS simulation for the same time snapshots as in the upper panel. The height of the TBFs is proportional to the squared norm of their amplitude.

Let us compare the QD dynamics to XFFMS, which would be exact in the limit of a large number of TBFs. At the early stage of the photoexcitation dynamics, we observe a perfect agreement between the position, shape, and width of the XFFMS nuclear wavepacket with the one obtained using exact QD (Fig. 3, upper panel, 1.2 fs). At later time, the nuclear wavepacket is reasonably well described, even if some discrepancies can be observed at 9.7 and 19.4 fs for the $S_1$ wavepacket due to the limited number of TBFs on $S_1$. The XFFMS dynamics indeed uses a maximum of four TBFs to describe the nuclear dynamics in $S_1$, which is not sufficient to accurately describe the nuclear wavepacket at these later times. The lower panel of Fig. 3 shows schematically the positions of the TBFs that generate the XFFMS nuclear wavepackets. Despite the differences in the shape of the nuclear wavepackets, the time-dependent population of the $S_1$ electronic state, computed as $n_{S_1}(t) = \sum_i^{N_{S_1}(t)} \sum_{i'}^{N_{S_1}(t)} C_i^{(S_1)}(t)^* C_i^{(S_1)}(t) S_{ii'}^{S_1 S_1}(t)$, is in excellent agreement with the exact QD simulation (Fig. 4, black line). What happens now if one applies the IFGA and approximates the matrix elements with a SPA of order 0, *i.e.*, if one employs XFAIMS for this photoexcitation process? Interestingly, XFAIMS also leads to a perfect agreement with both the XFAIMS and the QD dynamics for the population transfer to $S_1$ (Fig. 4, blue line). This agreement is, however, not surprising, considering that we apply a very short laser pulse to the molecule, as observed in the context of trajectory surface hopping.[94]

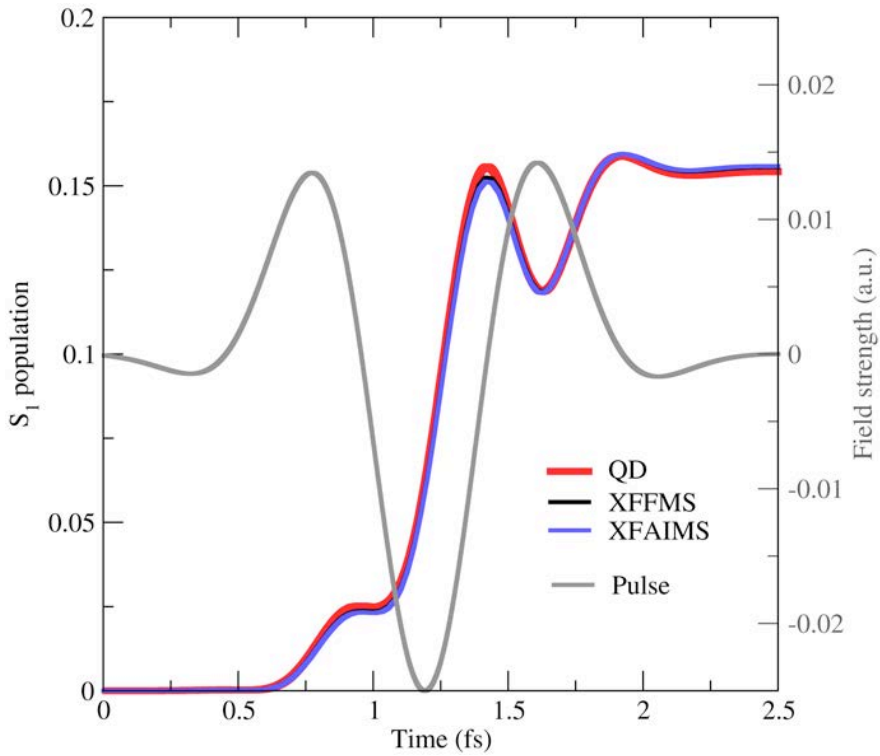

Figure 4: Time-dependent population of the $S_1$ state as obtained with XFFMS, XFAIMS, and with the numerically-exact QD method. The pulse ($f_0$=0.025 a.u., FWHM=35.25 a.u., $\omega$=0.127 a.u., CEP=$\pi$, polarization along the molecular axis) is depicted in gray.

Comparing XFFMS and XFAIMS (or any other nonadiabatic methods) with QD results using only the population as a metric is, however, not sufficient to evaluate the quality of a method's approximations. The (adiabatic) population only requires a scalar product of the nuclear wavefunction on the considered electronic state with itself, making it a rather forgiving observable with respect to different approximations. In the following, we will make use of a different observable, the time-dependent molecular dipole moment, to monitor the effect of the approximations in XFAIMS. This quantity is defined by

$$\langle \Psi | \underline{\mu} | \Psi \rangle_{\mathbf{r},\mathbf{R}} = \sum_{I}^{\infty} \sum_{J}^{\infty} \left[ \langle \Omega_I | \langle \Phi_I | \underline{\mu}^e | \Phi_J \rangle_{\mathbf{r}} | \Omega_J \rangle_{\mathbf{R}} + \langle \Omega_I | \underline{\mu}^n | \Omega_J \rangle_{\mathbf{R}} \delta_{IJ} \right], \quad (11)$$

based on the definition of the dipole moment operator given above. Within the representation of the nuclear wavefunctions proposed by XFFMS, the same quantity becomes

$$\langle\Psi|\underline{\mu}|\Psi\rangle_{\mathbf{r},\mathbf{R}} = \sum_{I}^{\infty}\sum_{J}^{\infty}\left[\sum_{k}^{N_I(t)}\sum_{i}^{N_J(t)}\left(C_k^{(I)}(t)\right)^*C_i^{(J)}(t)\left[\langle\chi_k^{(I)}|\langle\Phi_I|\underline{\mu}^e|\Phi_J\rangle_{\mathbf{r}}|\chi_i^{(J)}\rangle_{\mathbf{R}} + \langle\chi_k^{(I)}|\underline{\mu}^n|\chi_i^{(J)}\rangle_{\mathbf{R}}\delta_{IJ}\right]\right]. \quad (12)$$

The time-dependent dipole moment can be decomposed into three terms: (*i*) an electronic contribution combining couplings between TBFs in the same state and mediated by the electronic dipole moment; (*ii*) interference terms triggered by transition dipole moments and coming from the coupling between TBFs in different electronic states (importantly, this term is modulated by the strength of the transition dipole moments and the overlap between the TBFs on the two PECs, see Eq.(12)); (*iii*) the nuclear contribution to the dipole moment, which is diagonal in the adiabatic representation. Hence, the time-dependent dipole moment is composed of rather complex matrix elements between nuclear TBFs evolving on the same and on different electronic states, meaning that such observable will be sensitive to variations in both phase and spatial localization of the two different nuclear wavepackets. Therefore, this quantity constitutes a challenge for any approximated nuclear dynamics methods aiming at describing processes like interferences or decoherences adequately, and will be a central quantity for our analysis of XFFMS and XFAIMS approximations.

The time-dependent dipole moment obtained from the QD simulation rapidly oscillates, with a decrease of its overall amplitude as the nuclear wavepacket on $S_1$ leaves the Franck-Condon region (Fig. 5, red line). These oscillations originate from the interference between the electronic state $S_0$ and $S_1$ state and the beating is inversely proportional to the $S_0$-$S_1$ excitation energy. The time-dependent dipole moment computed with XFFMS agrees very well with the QD simulation, but we note a slight dephasing between 10 and 15 fs (Fig. 5, black line). Based on our previous discussion on the derivation of (XF)FMS, this slight dephasing of the dipole moment in XFFMS – and therefore the slight deviation from the numerically exact results – should find its root in the use of a small number of TBFs ("Truncated Basis" in Fig. 1). To demonstrate this, we carried out an additional XFFMS simulation in which we artificially increased the number of basis functions on $S_1$ by spreading 73 fixed Gaussian functions that cover the whole region visited by the nuclear wavepacket on $S_1$ (ranging from 2.83 to 10.03 Bohr). The fixed Gaussians have the very same definition as the TBFs, except that their momentum is always zero, their position do not move over time (we still use 9 moving TBFs on $S_0$), and no spawning events are required. This Fixed Gaussian (FG) dynamics constitutes an intermediate level between the numerically exact QD (complete fixed grid) and XFFMS (truncated travelling basis). As expected, the time-dependent dipole moment obtained with the FG method is in excellent agreement with the exact QD simulation, even when the nuclear wavepacket on $S_1$ leaves the FC region (Fig. 5, orange line). Nevertheless, it should be stressed that the almost quantitative result obtained with XFFMS only uses a total of 13 TBFs, with just four of them evolving on $S_1$, leading to a drastic reduction of the computational cost as compared to the FG simulation where 82 TBFs in total were used.

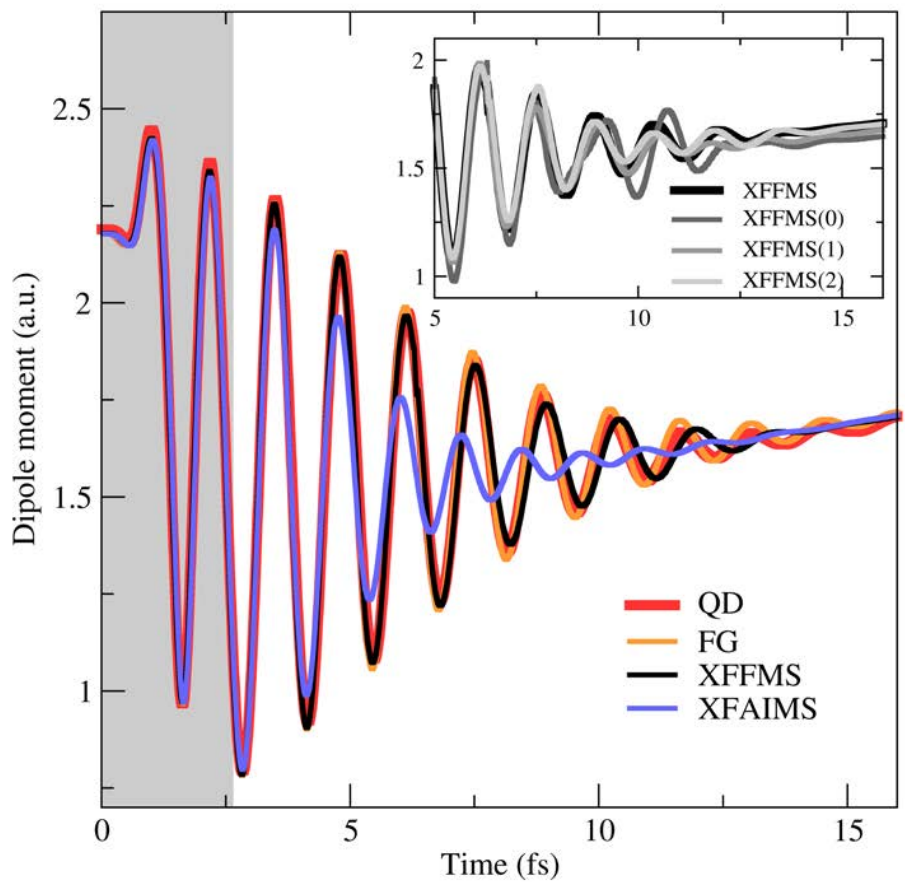

Figure 5: Time-dependent dipole moment in the first 16 fs following photoexcitation as computed with XFFMS, XFAIMS, FG, and exact QD. The grey area represents the time window in which the laser pulse is applied to the molecule. The effect of the SPA on XFFMS is shown in the inset for the dipole moment between 5 and 16 fs, when the nuclear wavepacket on $S_1$ is leaving the FC region.

Now that we showed that XFFMS can be converged towards the exact QD dynamics in the limit of a large basis set, let us focus on the other approximation that can be applied to XFFMS: the SPA on Hamiltonian matrix elements. As described above, XFFMS would correspond to an infinite number of terms in the Taylor expansion or the knowledge of the electronic quantities over the entire configuration space visited by the nuclear wavepackets (Fig. 1). As expected, the time-dependent dipole moment is sensitive to the approximations applied to XFFMS as it reflects the interferences between the electronic states *via* the motion of the nuclear wavepackets and the overlap between the two wavepackets. The XFFMS-SPA0 dipole moment starts to dephase after 7 fs, when the $S_1$ wavepacket leaves the FC region (see inset in Fig. 5). It is at that particular time that the overlap between the $S_0$ and $S_1$ wavepackets – and therefore between the corresponding TBFs – decreases. As the TBFs are spatially separated, approximating the electronic structure quantities in the Hamilonian matrix elements by single (constant) values evaluated at the centroid position of their product does not approximate well the integral over the full span of the two TBFs. Employing the SPA-1, which requires the additional calculation of the first derivative of the electronic energy and the (transition) dipole moments (see Eq. (9)), in the dynamics already leads to a dramatic improvement in the description of the time-dependent dipole moment (see inset in Fig. 5). The time-dependent dipole moment now beats in phase but the amplitude is still off. This result can be improved by moving to the $2^{nd}$ or higher-order SPA that rapidly leads to a perfect agreement with XFFMS (*i.e.*, with full numerical integration for computing the Hamiltonian matrix elements).

We now move to the second approximation used in XFAIMS: the independent first generation. We first note that the IFGA is justified for multidimensional systems, but applying it to our one-dimensional systems constitutes its worst-case scenario. Comparing the exact QD simulation with XFAIMS, in which the IFGA is applied and the Hamiltonian matrix elements are computed using the $0^{th}$ order SPA, shows an excellent agreement for the $S_1$ population (Fig. 4). The situation is slightly worse in the case of the time-dependent dipole moment, where the coupling between all the TBFs becomes essential. A typical XFAIMS run consists in one TBF in $S_0$ (whose initial conditions were sampled from a Wigner distribution) that will eventually spawn another TBF in $S_1$. 100 XFAIMS runs are then sampled and incoherently averaged to produce the result presented here. Hence, the IFGA dramatically reduces the number of coupled TBFs: in our simulations, we move from 13 coupled TBFs in XFFMS to 2 per run in XFAIMS. While the time evolution of the dipole moment is still qualitatively well described by XFAIMS, the amplitude of its oscillations decreases faster than for the QD simulation, and a dephasing is observed at later time (Fig. 5). This difference can be explained by the fact that the (uncoupled) TBFs rapidly leave the FC region, and therefore do not offer a proper support to describe the slower dynamics component observed for the $S_1$ nuclear wavepacket in the QD simulation. Employing only two coupled TBFs in XFAIMS is enough to describe the population transfer with a short laser pulse, but such a reduced number of uncoupled TBFs might become problematic if one wants to describe the overall nuclear dynamics at later time. This situation could be improved by using a different set of initial conditions with zero initial momentum, for example, such that the $S_1$ TBFs remain for a longer period in the FC region – hence increasing the oscillation amplitude of the time-dependent dipole moment.

### *IVb. Step B: Return of the $S_1$ nuclear wavepacket in the Franck-Condon region*

The previous section was dedicated to the early dynamics following photoexcitation, when the nuclear wavepacket on $S_1$ starts to leave the FC region, but what happens at later times? The nuclear wavepacket moves towards longer LiH distance until it reflects (after 35 fs) and comes back into the FC region (Fig. 6, left). The time-dependent dipole moment mirrors this dynamics as it slowly increases from 1.25 a.u. to 2.0 a.u. during the first 40 fs of dynamics before decreasing again – the $S_1$ electronic dipole moment indeed increases with the LiH bond length. When the $S_1$ nuclear wavepacket comes back into the FC region, between 60 and 80 fs, we observe oscillations in the time-dependent dipole moment caused by interferences between the nuclear wavepackets in electronic states $S_0$ and $S_1$ (Fig. 6). XFAIMS qualitatively reproduces this trend in the time-dependent dipole moment (XFFMS does it almost quantitatively), attesting from the proper description of the nuclear wavepacket dynamics in each electronic state.

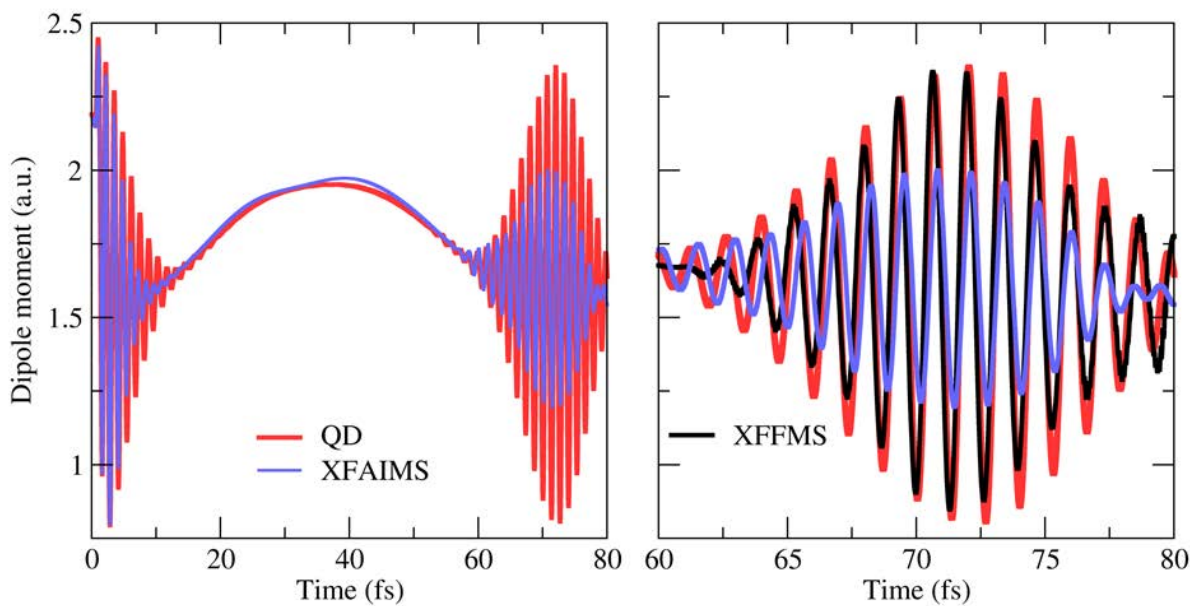

Figure 6: Time-dependent dipole moment induced by the UV pulse of Fig. 4 computed with XFAIMS and with the QD method. The dipole moment is showed in a time-window from 0 to 80 fs (left) and zoomed from 60 to 80 fs (right). For the right panel the dipole moment computed with XFFMS is also shown.

*IVc. Step C: Pump-probe experiment*

An interesting way to probe the return of the $S_1$ nuclear wavepacket into the FC region consists in applying a second laser pulse, a *probe*, at a variable time delay after the original *pump* pulse (Fig. 2c). Depending on the phase between the two nuclear wavepackets and their overlap at the time of the second pulse, the population will either be excited from $S_0$ to $S_1$ or de-excited from $S_1$ to $S_0$. We therefore expect an oscillation of the resulting $S_1$ population at the end of the second pulse as a function of the pump-probe delay (Fig. 7). Such pump-probe experiment is rather challenging to model for methods outside quantum dynamics, because it requires to accurately describe the photoexcitation and the nuclear wavepackets propagation during more than 60 fs. Due to the coupling between TBFs and the proper treatment of coherence and decoherence effects, XFFMS and XFAIMS are expected to be suitable to model such experiments, unlike trajectory surface hopping that might suffer from its inherent *independent trajectory approximation*.[25, 94, 95] XFAIMS recovers the proper population beating (Fig. 7), even if the oscillations of the $S_1$ population are weaker than those obtained with the exact QD simulation. As previously noted, this effect is due to the width of the approximate nuclear wavepackets within the IFGA: the $S_0$-$S_1$ overlap is underestimated by a factor of two when the $S_1$ nuclear wavepacket returns into the FC region, explaining why the population transfer is smaller. This problem can be circumvented by releasing the IFGA, as demonstrated by the result produced with XFFMS (using this time 9 TBFs on $S_0$ and 9 on $S_1$). The $S_1$ population after the second pulse is now in very good agreement with the grid simulation.

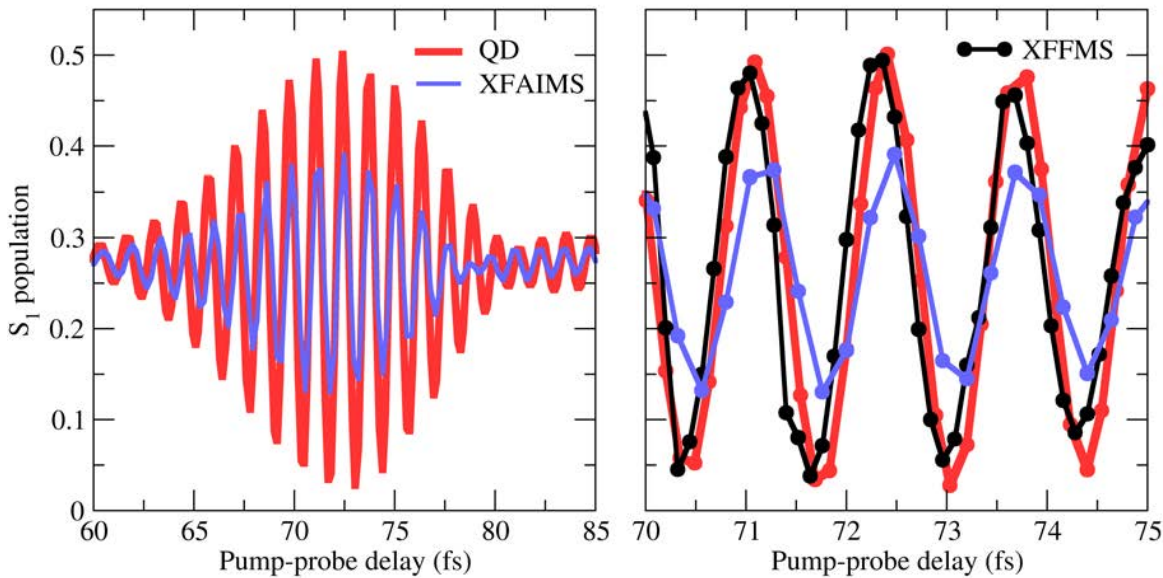

Figure 7: Evolution of the $S_1$ population after the probe by the second pulse as a function of the pump-probe delay between 60 and 85fs (left) and 70 and 75fs (right) computed in XFAIMS and with the QD method. In the right panel the population computed with XFFMS is also shown.

### IVd. Discussion on AIMS and possible improvements

The previous sections proposed an analysis of the different approximations bridging the (XF)FMS framework to (XF)AIMS. We generalize here our findings to AIMS and discuss potential improvements of the method. AIMS employs a SPA-0 which, as described above, appears to capture the qualitative features of the time-dependent dipole moment adequately. This approximation nevertheless rapidly runs out of steam for matrix elements between TBFs that are moving apart (inset of Fig. 5), and we observed that moving to the SPA-1 leads to a substantial improvement of the result. However, the SPA-1 implies the additional calculation of derivatives of electronic structure quantities with respect to nuclear coordinates at the centroid position. However, such quantities are naturally computed at the center of each TBFs – nuclear gradients are indeed required for the classical propagation of each TBFs. Based on this observation, Shalashilin and Martínez proposed to replace the SPA-1 by a first-order bra-ket averaged Taylor (BAT-1) expansion,[76] where matrix elements are given by:

$$
\begin{aligned}
\langle \chi_k^{(I)} | \vartheta | \chi_i^{(J)} \rangle_{\mathbf{R}} &\approx \frac{1}{2} \left( \vartheta(\bar{\mathbf{R}}_k^{(I)}) + \vartheta(\bar{\mathbf{R}}_i^{(J)}) \right) \langle \chi_k^{(I)} | \chi_i^{(J)} \rangle_{\mathbf{R}} \\
&+ \frac{1}{2} \left( \sum_\rho^{3N} \frac{\partial \vartheta(\mathbf{R})}{\partial R_\rho} \bigg|_{R_\rho = \bar{R}_{\rho,k}^{(I)}} \langle \chi_k^{(I)} | (R_\rho - \bar{R}_{\rho,k}^{(I)}) | \chi_i^{(J)} \rangle_{\mathbf{R}} \right) \\
&+ \frac{1}{2} \left( \sum_\rho^{3N} \frac{\partial \vartheta(\mathbf{R})}{\partial R_\rho} \bigg|_{R_\rho = \bar{R}_{\rho,i}^{(J)}} \langle \chi_k^{(I)} | (R_\rho - \bar{R}_{\rho,i}^{(J)}) | \chi_i^{(J)} \rangle_{\mathbf{R}} \right)
\end{aligned}
\quad (13)
$$

The BAT-1 does not require any additional electronic-structure calculation, neither at the centroid nor at the TBF position.[38, 76] Armed with our exact model, we tested the BAT-1 approximation and compared it with the SPA-1, as this strategy was only tested empirically on molecular systems. The BAT-1 reproduces accurately the SPA-1 dynamics (Fig. 8), hence providing a clear improvement over the SPA-0 at no extra cost, and its use should be strongly advised for any methods based on travelling Gaussians. A question that could also arise at this point is: shall we strictly employ the same order of the SPA for intra and interstate couplings? Let us make two observations: (*i*) the spawning algorithm ensures that TBFs are strongly overlapping in regions where the coupling between states is maximum, and (*ii*) our present laser pulse – as well as NACVs more generally – are believed to behave ideally as *localized couplings*. Based on these two observations, one could suggest that a lower order SPA could be employed for interstate coupling matrix elements, while a higher one should be used for the intrastate ones. We performed this test for our model system and observed an almost perfect agreement between the time-dependent dipole moment simulated with the SPA-1 for all matrix elements, and the one obtained by employing a SPA-0 for interstate couplings and a SPA-1 for intrastate coupling (inset of Fig. 8). This result is particularly interesting for AIMS, as employing a SPA-1 (or even a BAT-1) for its matrix elements would imply the calculation of higher-order order terms related to the NACVs that are currently not implemented in electronic structure codes. One could therefore suggest that, if higher accuracy is needed in an AIMS simulation, the IFGA can be released and the SPA-1 (or BAT-1) applied to intrastate couplings (keeping a SPA-0 for interstates ones).

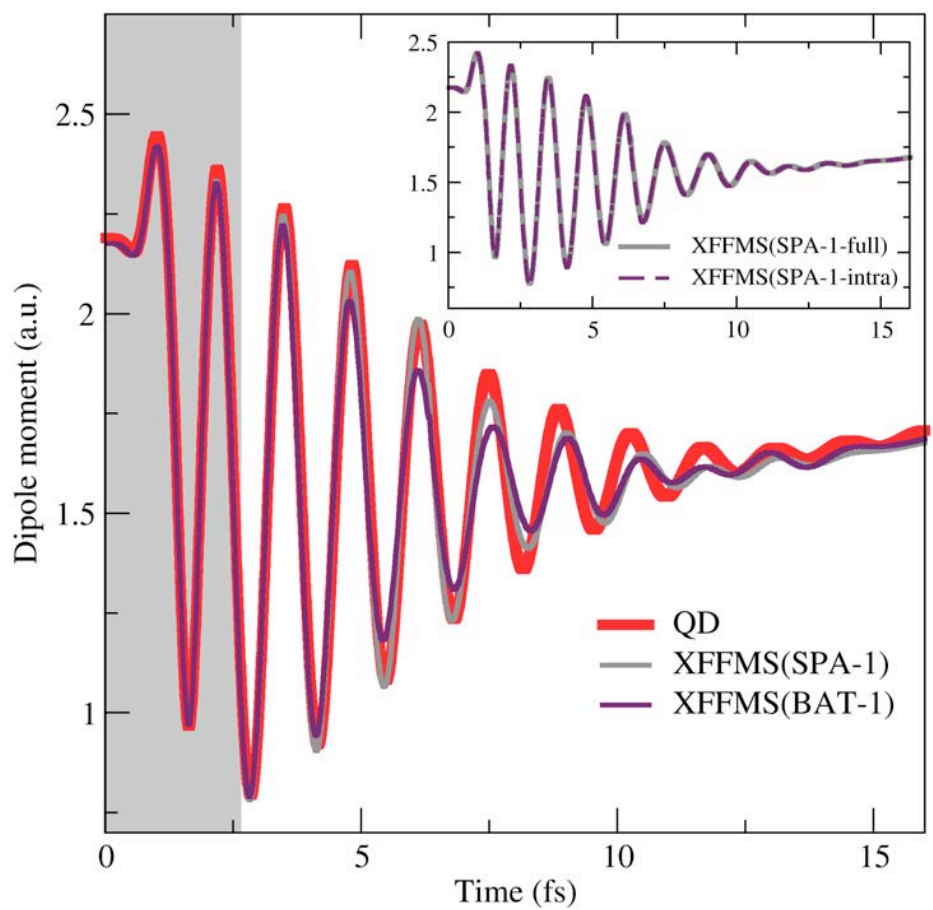

Figure 8: Time-dependent dipole moment in the first 16 fs computed with XFFMS using the SPA-1 (grey line), the BAT-1 (palatinate line), and QD (red line). Inset: resulting time-dependent dipole moment when the SPA-1 is only applied to intrastate coupling terms (SPA-1-intra, dashed palatinate line), as compared to the full SPA-1 (grey line).

## V. Conclusion

In this work, we showed numerically how the (XF)FMS framework can be converged towards numerically exact QD. More importantly, we analyzed the approximations applied to the (XF)FMS equations to generate the (XF)AIMS method, which is commonly employed to simulate the excited-state dynamics of molecules in their full nuclear configuration space. Comparing the time evolution of the dipole moment of LiH after photoexcitation for the different methods offers a rather stringent metric to test the quality of the (XF)AIMS approximations – more challenging than simply monitoring the electronic state populations over time, for example. Furthermore, we focused on three different steps following photoexcitation: (*i*) the early-stage dynamics, when the $S_1$ nuclear wavepacket leaves the FC region, (*ii*) the long-time dynamics, when the $S_1$ wavepacket returns into the FC region, and (*iii*) the resulting population after a second pulse is applied to the system after a given delay. Despite being used in its worst-case scenario, *i.e.*, for a one-dimensional system, the IFGA does not prevent (XF)AIMS from capturing the proper physics of the nuclear wavepacket dynamics in all three steps. As expected, the SPA offers a decent approximation to the Hamiltonian matrix elements, but its lower-order version, SPA-0, starts to break down when the nuclear wavepackets in different electronic states separate while still experiencing a non-zero overlap. For the system studied, the SPA-1 already corrects this deficiency substantially and can easily be approximated, as proposed in earlier work and validated here. This work highlights that (XF)AIMS, despite its approximations, adequately describes the physics of rather challenging excited-state processes such as interferences or pump-probe pulse sequences, and therefore constitutes a robust technique to treat the excited-state dynamics of molecular systems. Furthermore, releasing the IFGA and employing a SPA-1 or a BAT-1 (only on intrastate couplings or on all couplings) allows for a simple test of XFAIMS accuracy. Finally, this work paves the way for the development of intermediate schemes between (XF)FMS and (XF)AIMS that will allow to selectively tune the accuracy of Hamiltonian matrix elements when required.

**VI. Acknowledgments**

The authors thank Federica Agostini for helpful comments on the manuscript. B.M. gratefully acknowledges support from the Fonds National de la Recherche Scientifique (Belgium). Computational resources have been provided by the Consortium des Equipements de Calcul Intensif (CECI), funded by the Fonds de la Recherche Scientifique de Belgique (F.R.S.-FNRS) under Grant No. 2.5020.11. This work also made use of the facilities of the Hamilton HPC Service of Durham University.

**VII. References**

[1] J. C. Tully, J. Chem. Phys. **137**, 22A301 (2012).

[2] D. R. Yarkony, Chem. Rev. **112**, 481 (2012).

[3] T. Yonehara, K. Hanasaki, and K. Takatsuka, Chem. Rev. **112**, 499 (2012).

[4] M. H. Beck, A. Jäckle, G. A. Worth, and H. D. Meyer, Phys. Rep. **324**, 1 (2000).

[5] G. A. Worth, H. D. Meyer, and L. S. Cederbaum, in *Conical Intersections: Electronic Structure, Dynamics & Spectroscopy*, edited by W. Domcke, D. R. Yarkony, and H. Köppel (World Scientific Publishing Co. Pte. Ltd., 2004), pp. 583.

[6] J. C. Tully, and R. K. Preston, J. Chem. Phys. **55**, 562 (1971).

[7] J. C. Tully, J. Chem. Phys. **93**, 1061 (1990).

[8] M. F. Herman, J. Chem. Phys. **81**, 754 (1984).

[9] X. Sun, and W. H. Miller, J. Chem. Phys. **106**, 6346 (1997).

[10] R. Kapral, and G. Ciccotti, J. Chem. Phys. **110**, 8919 (1999).

[11] S. Nielsen, R. Kapral, and G. Ciccotti, J. Stat. Phys. **101**, 225 (2000).

[12] S. Nielsen, R. Kapral, and G. Ciccotti, J. Chem. Phys. **112**, 6543 (2000).

[13] R. Kapral, Ann. Rev. Phys. Chem. **57**, 129 (2006).

[14] S. J. Cotton, and W. H. Miller, J. Chem. Phys. **139**, 234112 (2013).

[15] S. J. Cotton, and W. H. Miller, J. Phys. Chem. A **117**, 7190 (2013).

[16] S. J. Cotton, and W. H. Miller, J. Phys. Chem. A **119**, 12138 (2015).

[17] S. J. Cotton, and W. H. Miller, J. Chem. Phys. **145**, 144108 (2016).

[18] S. J. Cotton, and W. H. Miller, J. Chem. Theo. Comp. **12**, 983 (2016).

[19] S. Bonella, and D. F. Coker, J. Chem. Phys. **122**, 194102 (2005).

[20] R. E. Wyatt, C. L. Lopreore, and G. Parlant, J. Chem. Phys. **114**, 5113 (2001).

[21] C. L. Lopreore, and R. E. Wyatt, J. Chem. Phys. **116**, 1228 (2002).

[22] V. A. Rassolov, and S. Garashchuk, Phys. Rev. A **71**, 032511 (2005).

[23] B. Poirier, and G. Parlant, J. Phys. Chem. A **111**, 10400 (2007).

[24] B. F. E. Curchod, I. Tavernelli, and U. Rothlisberger, Phys. Chem. Chem. Phys. **13**, 3231 (2011).

[25] B. F. E. Curchod, and I. Tavernelli, J. Chem. Phys. **138**, 184112 (2013).

[26] F. Agostini, A. Abedi, Y. Suzuki, and E. K. U. Gross, Mol. Phys. **111**, 3625 (2013).

[27] A. Abedi, F. Agostini, and E. K. U. Gross, Europhys. Lett. **106**, 33001 (2014).

[28] S. K. Min, F. Agostini, and E. K. U. Gross, Physical Review Letters **115**, 2015).

[29] F. Agostini, S. K. Min, A. Abedi, and E. K. U. Gross, J. Chem. Theo. Comp. **12**, 2127 (2016).

[30] S. K. Min, F. Agostini, I. Tavernelli, and E. K. U. Gross, J. Phys. Chem. Lett. **8**, 3048 (2017).

[31] E. J. Heller, J. Chem. Phys. **62**, 1544 (1975).

[32] E. J. Heller, Acc. Chem. Res. **14**, 368 (1981).

[33] E. J. Heller, J. Chem. Phys. **75**, 2923 (1981).

[34] G. Worth, M. Robb, and I. Burghardt, Faraday Disc. **127**, 307 (2004).

[35] G. W. Richings, I. Polyak, K. E. Spinlove, G. A. Worth, I. Burghardt, and B. Lasorne, Int. Rev. Phys. Chem. **34**, 269 (2015).

[36] D. V. Shalashilin, J. Chem. Phys. **130**, 244101 (2009).

[37] D. V. Shalashilin, J. Chem. Phys. **132**, 244111 (2010).

[38] D. V. Makhov, C. Symonds, S. Fernandez-Alberti, and D. V. Shalashilin, Chem. Phys., doi:10.1016/j.chemphys.2017.04.003 (2017).

[39] T. J. Martínez, M. Ben-Nun, and R. D. Levine, J. Phys. Chem. **100**, 7884 (1996).

[40] T. J. Martínez, and R. D. Levine, J. Chem. Soc., Faraday Trans. **93**, 941 (1997).

[41] T. J. Martínez, and R. D. Levine, J. Chem. Phys. **105**, 6334 (1996).

[42] M. Ben-Nun, and T. J. Martínez, J. Chem. Phys. **108**, 7244 (1998).

[43] M. Ben-Nun, and T. J. Martínez, Adv. Chem. Phys. **121**, 439 (2002).

[44] M. Ben-Nun, J. Quenneville, and T. J. Martínez, J. Phys. Chem. A **104**, 5161 (2000).

[45] D. M. Leitner, J. Quenneville, B. Levine, T. J. Martínez, and P. G. Wolynes, J. Phys. Chem. **107**, 10706 (2003).

[46] B. Levine, and T. J. Martínez, in *Quantum Dynamics and Conical Intersections*, edited by G. A. Worth, and S. C. Allthorpe (CCP6, Daresbury, 2004), p. 65.

[47] J. D. Coe, and T. J. Martínez, J. Phys. Chem. A **110**, 618 (2006).

[48] H. R. Hudock, B. G. Levine, A. L. Thompson, H. Satzger, D. Townsend, N. Gador, S. Ullrich, A. Stolow, and T. J. Martínez, J. Phys. Chem. A **111**, 8500 (2007).

[49] B. G. Levine, and T. J. Martinez, Ann. Rev. Phys. Chem. **58**, 613 (2007).

[50] H. Tao, T. K. Allison, T. W. Wright, A. M. Stooke, C. Khurmi, J. van Tilborg, Y. Liu, R. W. Falcone, A. Belkacem, and T. J. Martínez, J. Chem. Phys. **134**, 244306 (2011).

[51] T. K. Allison, H. Tao, W. J. Glover, T. W. Wright, A. M. Stooke, C. Khurmi, J. van Tilborg, Y. Liu, R. W. Falcone, T. J. Martínez, and A. Belkacem, J. Chem. Phys. **136**, 124317 (2012).

[52] C. Punwong, J. Owens, and T. J. Martinez, J. Phys. Chem. B **119**, 704 (2015).

[53] B. Mignolet, B. F. E. Curchod, and T. J. Martinez, Angew. Chem. Int. Ed. **55**, 14993 (2016).

[54] J. W. Snyder, B. F. E. Curchod, and T. J. Martinez, J. Phys. Chem. Lett. **7**, 2444 (2016).

[55] B. F. E. Curchod, A. Sisto, and T. J. Martinez, J. Phys. Chem. A **121**, 265 (2017).

[56] S. Pijeau, D. Foster, and E. G. Hohenstein, J. Phys. Chem. A **121**, 6377 (2017).

[57] S. Pijeau, D. Foster, and E. G. Hohenstein, J. Phys. Chem. A **121**, 4595 (2017).

[58] M. Sulc, H. Hernandez, T. J. Martínez, and J. Vanicek, J. Chem. Phys. **139**, 2013).

[59] M. D. Hack, A. M. Wensmann, D. G. Truhlar, M. Ben-Nun, and T. J. Martínez, J. Chem. Phys. **115**, 1172 (2001).

[60] J. P. Alborzpour, D. P. Tew, and S. Habershon, J. Chem. Phys. **145**, 174112 (2016).

[61] J. E. Subotnik, and N. Shenvi, J. Chem. Phys. **134**, 024105 (2011).

[62] M. Born, and K. Huang, *Dynamical Theory of Crystal Lattices* (Clarendon, Oxford, 1954),

[63] W. Domcke, D. Yarkony, and H. Köppel, *Conical Intersections: Electronic Structure, Dynamics & Spectroscopy* (World Scientific Pub Co Inc, 2004), Vol. 15.

[64] A. Abedi, N. T. Maitra, and E. K. U. Gross, Phys. Rev. Lett. **105**, 123002 (2010).

[65] A. Abedi, N. T. Maitra, and E. K. U. Gross, J. Chem. Phys. **137**, 22A530 (2012).

[66] A. Abedi, F. Agostini, Y. Suzuki, and E. K. U. Gross, Phys. Rev. Lett **110**, 263001 (2013).

[67] H.-D. Meyer, F. Gatti, and G. A. Worth, *Multidimensional Quantum Dynamics: MCTDH Theory and Applications* (Wiley-VCH Verlag GmbH & Co. KGaA, 2009),

[68] F. Gatti, and B. Lasorne, Phys Chem Action, 271 (2014).

[69] D. J. Tannor, *Introduction to quantum mechanics, a time-dependent perspective* (University Science Books, Sausalito, California, 2007),

[70] G. A. Worth, H.-D. Meyer, H. Köppel, L. S. Cederbaum, and I. Burghardt, Int. Rev. Phys. Chem. **27**, 569 (2008).

[71] K. Saita, and D. V. Shalashilin, J. Chem. Phys. **137**, 22a506, 8 (2012).

[72] K. Saita, M. G. D. Nix, and D. V. Shalashilin, Phys. Chem. Chem. Phys. **15**, 16227 (2013).

[73] G. A. Worth, and I. Burghardt, Chem. Phys. Lett. **368**, 502 (2003).

[74] B. Lasorne, M. J. Bearpark, M. A. Robb, and G. A. Worth, Chem. Phys. Lett. **432**, 604 (2006).

[75] B. Lasorne, M. A. Robb, and G. A. Worth, Phys. Chem. Chem. Phys. **9**, 3210 (2007).

[76] D. V. Makhov, W. J. Glover, T. J. Martinez, and D. V. Shalashilin, J. Chem. Phys. **141**, 054110, 11 (2014).

[77] D. V. Makhov, K. Saita, T. J. Martinez, and D. V. Shalashilin, Phys. Chem. Chem. Phys. **17**, 3316 (2015).

[78] L. Joubert-Doriol, J. Sivasubramanium, I. G. Ryabinkin, and A. F. Izmaylov, J. Phys. Chem. Lett. **8**, 452 (2017).

[79] G. A. Meek, and B. G. Levine, J. Chem. Phys. **145**, 184103 (2016).

[80] R. Gherib, L. Y. Ye, I. G. Ryabinkin, and A. F. Izmaylov, J. Chem. Phys. **144**, 154103 (2016).

[81] G. A. Meek, and B. G. Levine, J. Chem. Phys. **144**, 184109 (2016).

[82] F. G. Eich, and F. Agostini, J. Chem. Phys. **145**, 054110 (2016).

[83] S. Yang, J. D. Coe, B. Kaduk, and T. J. Martínez, J. Chem. Phys. **130**, 2009).

[84] S. Yang, and T. J. Martínez, in *Conical Intersections: Theory, Computation and Experiment*, edited by W. Domcke, D. R. Yarkony, and H. Köppel (World Scientific Publishing Co. Pte. Ltd., 2011), pp. 347.

[85] B. Mignolet, B. F. E. Curchod, and T. J. Martinez, J. Chem. Phys. **145**, 2016).

[86] A. Nikodem, and F. Remacle, J Phys Conf Ser **635**, 112066 (2015).

[87] A. Nikodem, R. D. Levine, and F. Remacle, J. Phys. Chem. A **120**, 3343 (2016).

[88] F. Remacle, M. Nest, and R. D. Levine, Phys. Rev. Lett. **99**, 183902 (2007).

[89] B. Mignolet, A. Gijsbertsen, M. J. J. Vrakking, R. D. Levine, and F. Remacle, Phys. Chem. Chem. Phys. **13**, 8331 (2011).

[90] A. Nikodem, R. D. Levine, and F. Remacle, Phys. Rev. A **95**, 053404 (2017).

[91] B. G. Levine, J. D. Coe, A. M. Virshup, and T. J. Martínez, Chemical Physics **347**, 3 (2008).

[92] H.-J. Werner, P. J. Knowles, F. R. Manby, M. Schuetz, P. Celani, G. Knizia, T. Korona, R. Lindh, A. Mitrushenkov, G. Rauhut, T. B. Adler, R. D. Amos, A. Bernhardsson, A. Berning, D. L. Cooper, M. J.

O. Deegan, A. J. Dobbyn, F. Eckert, E. Goll, C. Hampel, A. Hesselmann, G. Hetzer, T. Hrenar, G. Jansen, C. Koeppl, Y. Liu, A. W. Lloyd, R. A. Mata, A. J. May, S. J. McNicholas, W. Meyer, M. E. Mura, A. Nicklass, P. Palmieri, K. Pflueger, R. Pitzer, M. Reiher, T. Shiozaki, H. Stoll, A. J. Stone, R. Tarroni, T. Thorsteinsson, M. Wang, and A. Wolf, Cardiff, UK, 2012).

[93] A. L. Thompson, C. Punwong, and T. J. Martínez, Chem. Phys. **370**, 70 (2010).

[94] J. J. e. Bajo, G. Granucci, and M. Persico, J. Chem. Phys. **140**, 044113 (2014).

[95] J. E. Subotnik, A. Jain, B. Landry, A. Petit, W. J. Ouyang, and N. Bellonzi, Ann. Rev. Phys. Chem. **67**, 387 (2016).